\documentclass[atoms,article,accept,pdftex,moreauthors]{Definitions/mdpi}

\firstpage{1} 
\pubvolume{1}
\issuenum{1}
\articlenumber{0}
\pubyear{2023}
\copyrightyear{2023}
\externaleditor{{Academic} Editor: Sultana N. Nahar, Guillermo Hinojosa}%MDPI please add academic editor
\datereceived{ 06 February 2023} 
\daterevised{ 08 March 2023} % Comment out if no revised date
\dateaccepted{ 13 March 2023  } 

\datepublished{ } 
\hreflink{https://doi.org/} % If needed use \linebreak
\Title{The Shapes of Stellar Spectra}

\TitleCitation{The Shapes of Stellar Spectra}

\Author{Carlos %MDPI: Please carefully check the accuracy of names and affiliations. There is no explanation of dagger and ddagger so we delete them. please confirm
 Allende Prieto $^{1,2}$\orcidA{0000-0002-0084-572X}}
\AuthorNames{Carlos Allende Prieto}

\AuthorCitation{Allende %MDPI: Please check all author names carefully.
 Prieto, C.}
\address{%
$^{1}$ \quad Instituto de Astrof\'{\i}sica de Canarias, V\'{\i}a L\'actea S/N, E-38205 La Laguna, Tenerife, %MDPI is it necessary? %Author Yes
Spain; callende@iac.es;\linebreak Tel.:  +34-922-605200\\
$^{2}$ \quad Universidad de La Laguna, Departamento %MDPI: please confirm the order of affiliation. %Author Ok
 de Astrof\'{\i}sica, E-38206 La Laguna, Tenerife, Spain} %Author Swap order of University - Dept. to match standard form

\abstract{Stellar atmospheres separate the hot and dense stellar interiors 
from the emptiness of space. Radiation escapes from the outermost layers of 
a star, carrying  direct physical information. Underneath the atmosphere, the very high opacity keeps radiation 
thermalized and resembling a black body with the local temperature. 
In the atmosphere the opacity drops,  and radiative energy leaks out, which is redistributed in wavelength according to the physical processes 
by which matter and radiation interact, in particular photoionization. 
In this article, I will evaluate the role of photoionization in shaping the
stellar energy distribution of stars. To that end, I employ simple, state-of-the-art 
plane-parallel model atmospheres and a spectral synthesis code, dissecting the 
effects of photoionization from different chemical elements and species, for
stars of different masses in the range of 0.3 to 2 M$_{\odot}$. I examine and interpret 
the changes in the observed spectral energy distributions of the stars as 
a function of the atmospheric parameters. The photoionization of
atomic hydrogen and H$^-$ are the most relevant contributors to the continuum
opacity in the optical and near-infrared regions, while heavier elements
become important in the ultraviolet region. In the spectra of the coolest stars (spectral types M
and later), the continuum shape from photoionization is no longer recognizable due to 
the accumulation of lines, mainly from molecules.  These facts  have been known 
for a long time, but the calculations presented provide an updated quantitative evaluation 
and insight into the role of photoionization on the structure of stellar atmospheres.
}

\keyword{stars; stellar atmospheres; opacity} 

\begin{document}

%%%%%%%%%%%%%%%%%%%%%%%%%%%%%%%%%%%%%%%%%%
%\setcounter{section}{-1} %% Remove this when starting to work on the template.
\section{Introduction}

Starlight escapes from the atmospheres of stars, which become optically thin at
different heights depending on the wavelength. The main 
effect causing photon absorption in stellar atmospheres is photoionization.
As a result, this process reshapes the spectral energy distribution of stars, 
which in deeper layers must resemble black bodies, and
dictates the details of how their colors change with surface temperature (or mass).
In %MDPI: is the bold necessary? 
%Author No, it was introduced to highlight the changes in response to the referees and I remove them throughout the paper now
 what follows, the discussion will be focused on stars with masses between
0.3 and about 2 M$_{\odot}$. %MDPI is this dot necessary? %Author I think so
There are of course stars with less (down to 
the minimum mass for hydrogen burning, about 0.07 M$_{\odot}$), and more 
mass (probably up to $\sim$ 100 M$_{\odot}$), but most stars are included in this
range.

Photoionization in stellar atmospheres is mainly associated with hydrogen atoms
and closely related ions, in particular H$^-$, which is a dominant source of
absorption for solar-type stars. This is not surprising since 
80\% of the stellar mass is usually hydrogen, and the remaining is mostly helium, with 
less than 2\% left for the rest of the elements. 
Hydrogen photoionization imprints a series
of discontinuities on stellar spectra, related to the various atomic levels: the 
Lyman ($n=1$), Balmer ($n=2$), Paschem ($n=3$), Brackett ($n=4$), 
Pfund ($n=5$), etc. series, visible at wavenumbers of $R_H/n^2$, where  
$R_H$ is the Rydberg
constant, or wavelengths of 912, 3646, 8204, 14580, 22790 \AA, etc.
%These jumps in flux are  pressure sensitive, and therefore useful to
%constrain the stellar gravitational acceleration, usually considered
%constant given the small extent of the atmosphere compared to the 
%stellar radius. 
In thermodynamical equilibrium, the Saha equation shows that the ionization
fraction for H is inversely proportional to the electron density, 
and therefore higher electron pressure, as present  on a main-sequence star, burning 
hydrogen in its core, compared to an evolved giant star, leads to less 
ionization of H and more pronounced photoionization jumps.

For most stars, there is a competition between the opacity from photoionization 
of atomic H and H$^-$ at optical and infrared wavelenghts, 
and depending on this, the H series jumps become more
or less prominent. In the infrared region, from $\lambda \sim$ 16,000 \AA\ onwards, free-free
(inverse bremsstrahlung) typically overcomes bound-free (photoionization) for H$^-$.
In the ultraviolet region, the situation becomes much more complex, since there are 
multiple, heavier elements that become important contributors to the opacity through
ionization; usually carbon, sodium, magnesium, aluminum, silicon, 
%potassium? 
and iron become very important at different wavelengths.

Reports from the 1970s to  the 1990s claimed a significant mismatch
between models and observations in the ultraviolet for the Sun, suggesting a missing opacity 
source. More recent studies (see, e.g., [26,17,11]) % MDPI \hl{} %MDPI: Please cite all references with reference numbers, and place the numbers in square brackets ("[ ]"), e.g., [1], [1-3], or [1,3]. Please refer to the following website for more information: http://www.mdpi.com/authors/references. Please make sure every references have citation in main text
%Author Ok
claimed to have identified the  source as  iron, 
at least in part associated with autoionizing lines missed in the earliest calculations.

Nevertheless, the far UV spectrum of solar-like stars is very hard to model, since the
accumulated opacity makes photons to escape from the upper atmosphere (chromosphere, 
transition region, and corona), rather than the much-better-understood and easier-to-model photosphere. In those high layers, the low density, high temperatures, and the  presence of magnetic fields complicate physical modeling. 
For moderately warmer stars, on the other hand, such
as B-type stars, UV light escapes from deeper, photospheric, layers, which are 
simpler.

This paper, devoted to the role of photoionization in shaping 
stellar spectra, will first describe the data sources and codes 
used in our calculations.  In Section \ref{opacity}, I 
dissect the various
contributors to the opacity in the atmospheres of the Sun and other types 
of stars. Section \ref{observations} will test the models against selected
high-quality spectrophotometric observations, giving us an idea of the realism of 
the models, as well as their limitations. Section \ref{diagnostics} examines 
the sensitivity of the stellar
continuum to changes in the main atmospheric parameters: surface effective
temperature, surface gravity, and metallicity\endnote{The usual convention in astronomy is used for this parameter, which prescribes that all the elements 
heavier than He are changed
in the same ratio relative to their solar abundances, and that ratio 
is quantified with the parameter
 [Fe/H] $= \log \left( \frac{ {\rm N}_{\rm Fe} }{ {\rm N}_{\rm H} } \right) - 
           \log \left( \frac{ {\rm N}_{\rm Fe} }{ {\rm N}_{\rm H} } \right)_{\odot} $, 
           where N$_X$ is the number density for the element $X$.}  %MDPI: footnote is not allowed we chagned it into note format please confirm
%Author Ok
(the fraction of heavy elements).
The paper closes with a short summary and conclusions 
in Section \ref{conclusions}.

%%%%%%%%%%%%%%%%%%%%%%%%%%%%%%%%%%%%%%%%%%
\section{Adopted Data}
\label{data}

In the calculations below, we adopt classical 1D plane-parallel model atmospheres 
from the MARCS [18,19] and Kurucz ([11] and 
updates) grids---see also [29] for the most recent 
incarnations. The theory of stellar atmospheres is laid out in detail, 
for example, in the textbook [22].

The sources of atomic data adopted for the opacity and synthetic spectra
computed are mostly those described in [1], with 
some updates. Photoionization cross-sections are from TOPBASE 
[31]
for all the elements considered but iron, which are from [6]
 and [30]. Bound-free  and free-free absorption was also considered 
for H$^-$, H$_2^+$, He$^-$, CH, OH, H$_2^-$, as well as collisionally
induced opacity from H$_2$-H$_2$, H$_2$-He, H$_2$-H, and H-He. 
Rayleigh scattering on atomic hydrogen and helium, H$_2$, and the wings
of Lyman alpha were also included.

Atomic line data are from the most recent files from
Kurucz's website\endnote{\url{kurucz.harvard.edu} accessed on 25 October 2022, file gfall08oct17.dat} %MDPI: Please add the access date (Format: Date Month Year). e.g., (accessed on 1 January 2020).%Author done
updated with damping constants from [4,5]. 
Kurucz's website (and references therein) 
is also the source for the  molecular line data, including 
H$_2$, CH, C$_2$, CN, CO, NH, OH, MgH, SiH, SiO, AlO, CaH, CaO, 
CrH, FeH, MgO, NaH, SiH, and VO, with the exception of Exomol data employed for 
TiO [28] and H$_2$O [32].

The actual calculations were   performed with the latest version of 
the code Synspec [20,21]. The data
and the code used are bundled with the Python wrapper 
Synple\endnote{\url{github.com/callendeprieto/synple} accessed on 13 January 2023} %MDPI: Please add the access date (Format: Date Month Year). e.g., (accessed on 1 January 2020). %Author done
[20], version 1.2.

%%%%%%%%%%%%%%%%%%%%%%%%%%%%%%%%%%%%%%%%%%
\section{Atmospheric Opacity}
\label{opacity}

A star is a fairly independent entity, kept together by its own 
gravitational pull, and for most of its life producing its own 
energy through nuclear fusion in the core, where the temperature 
reaches tens of millions of degrees. Ionization is very high throughout
the stellar interior up to the surface, where 
the temperature drops under $T \sim$ 10,000 K, hydrogen becomes neutral, 
the electron density falls dramatically, and with it the material becomes 
transparent and radiation escapes. 

The opacity at the stellar surface shapes the stellar spectrum, which is
mainly due to the photoionization and inverse bremsstrahlung of the first 
few ionization stages of the most abundant elements, chiefly hydrogen 
and, in solar-type stars or cooler types,  H$^-$ and other molecules (see, 
e.g., [16]).

The top panel of Figure \ref{fig1} illustrates the run of temperature vs. density in 
the atmosphere of a solar-like star. The point marked with a circle
is just above the layers where the temperature of the star matches its
effective temperature ($T_{\rm eff}$, defined as that of a black body with the same 
radiative flux $F = \sigma T_{\rm eff}^4$, where $\sigma$
is the Stefan-Boltzmann constant, and from where the optical 
continuum is escaping. The bottom panel of the figure 
shows the total (blue) and photoionization
(plus bremsstrahlung) (orange) opacity at the chosen point ($T=5100$ K, 
$\rho = 1 \times 10^{-7}$  %MDPI: We correct scientific notations (e.g., "$8 \times 10^{3}$", not "8E3"). We corrected them. Please confirm.
%Author Ok
g cm$^{-3}$). 

As we mentioned in the introduction, the smooth curve that dominates 
the continuum opacity between 4000 and 
16,000 \AA\ is due to the photoionization of H$^-$, while the rising 
continuum curve at longer wavelengths is due to H$^-$ bremsstrahlung. 
In the UV region, photoionization from abundant neutral and singly ionized
elements is responsible for the ragged and rapidly increasing 
continuum opacity. For each ion, the total opacity is the sum of the 
contributions from multiple levels. 
The height over the continuum reached by many of the lines suggests
that line opacity is dominant, but most of these lines are narrow, with
an FWHM of a fraction of an \r{A}ngstrom, and the shape of the stellar 
spectrum is, at least in the optical and near-infrared regions, dominated by 
photoionization.

\begin{figure}[H]
\includegraphics[width=13.0 cm]{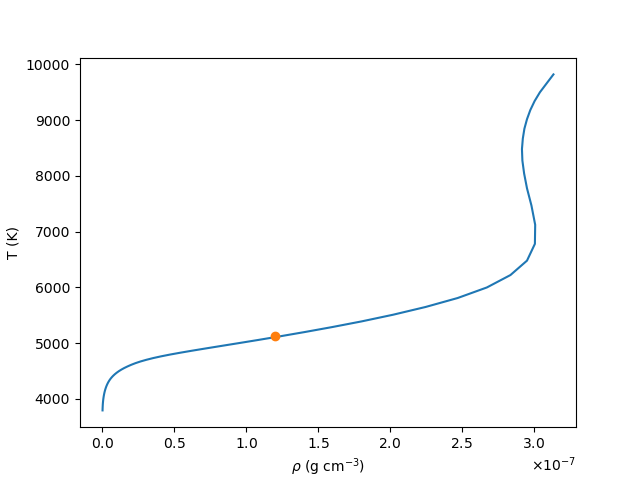}
\caption{\emph{Cont}. \label{fig1}}
\end{figure}

\begin{figure}[H]\ContinuedFloat
\includegraphics[width=13.0 cm]{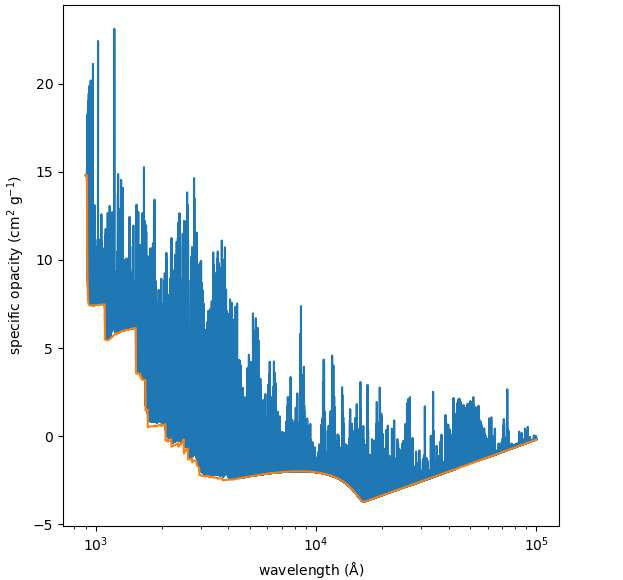}
\caption{(\textbf{Upper panel}) Relationship %MDPI: Please change the terms into scientific notation in the figure. (e.g., "$8 \times 10^{3}$", not "8E3"). %MDPI: Please use commas to separate thousands for numbers with five or more digits (not for four digits) in the picture. e.g., "10000" should be "10,000"
%Author Will try
 between temperature and 
density in the atmosphere of a solar-like star ($T_{\rm eff} = 5777$ K, 
$\log g  = 4.437$ with $g$ in cm s$^{-2}$, and the chemical abundances given 
in [2]) from a plane-parallel 
model in Local Thermodynamical Equilibrium. The indicated point is 
just slightly higher than optical depth unity, 
corresponding to $\rho  \simeq 1.2 \times 10^{-7}$ g cm$^{-3}$ and $T \simeq 5100$ K.
(\textbf{Lower panel}) Continuous (photoionization plus 
bremsstrahlung; orange)
 and total opacity for the point in the T-$\rho$ run marked 
 in the top panel. \label{fig1}}
\end{figure}

\begin{figure}[H]
\includegraphics[width=14.5 cm]{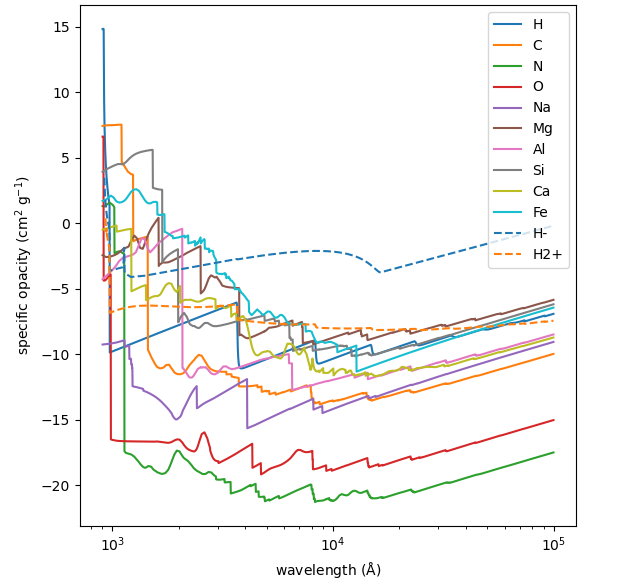}
\caption{Bound-free %MDPI: We moved all figures after it's first citaiton.
Author Ok
 and free-free opacity associated with individual
elements for $\rho  \simeq 1.2 \times 10^{-7}$ g cm$^{-3}$ and $T \simeq 5100$ K.
The  continuous opacity for H$^-$ and H$_2^+$ is also shown.
 \label{fig2}}
\end{figure}

Figure \ref{fig2} dissects the continuum opacity, for the same conditions
in the solar atmosphere adopted in Figure \ref{fig1}, into the various
atomic and molecular contributors. Neutral carbon, magnesium, aluminum, 
silicon, and iron show up as the most important absorbers in the UV region. However, 
while the important role of iron at $\lambda< 3000$ \AA\ and magnesium 
at $\lambda < 2500$ \AA\ is undeniable, the dramatic 
Increase in opacity, not only due to photoionization but also through
the accumulation of transitions at these wavelengths, shifts the layers
from which UV radiation escapes higher up, and the importance of
some of the ionization edges is hard to assess in 
practice. 

This picture may be incomplete, since the list of ions included
in the calculations is not exhaustive, and has in practice been limited
to those for which there are calculations available from the Opacity Project 
and the Iron Project. Furthermore, some of the opacity due to 
autoionization lines may be included twice as line transitions and 
resonances in the photoionization
cross sections. 

The agreement with observations (see Section \ref{observations}) suggests
the major contributors to opacity in stellar atmospheres have been 
identified, but there may be modestly or moderately important
 contributors missing. 
In a recent paper [8] it has been  pointed out 
that the free-free opacity cross-section 
for negative positronium ions (the equivalent of H$^-$ for a positronium 
instead of a H atom) would be larger than that of H$^-$ in the 
infrared for the solar atmosphere. 
Nonetheless, the lack of knowledge of the abundance of positrons in 
the solar atmosphere makes it hard to assess how much relevance such
a contribution would make to the total opacity.

\begin{figure}[H]
\includegraphics[width=12. cm]{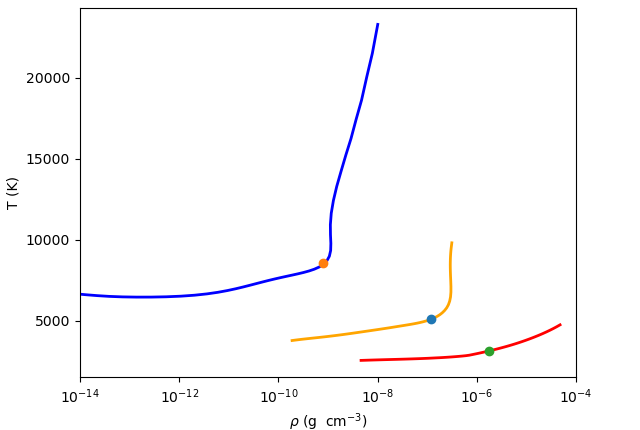}
\includegraphics[width=12. cm]{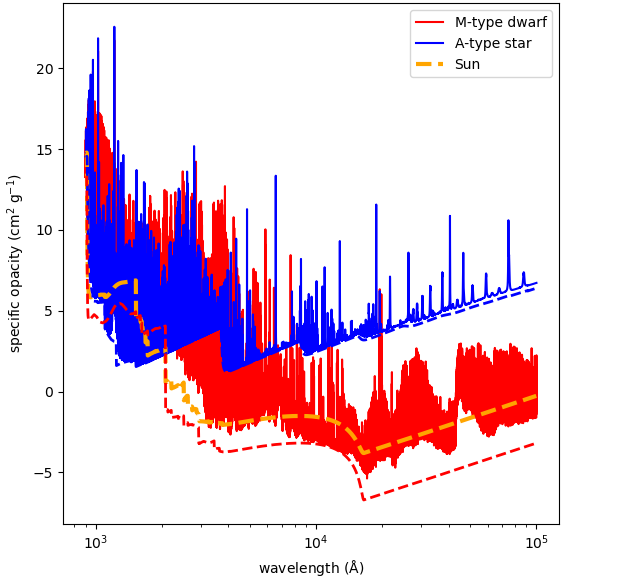}
\caption{Similar %MDPI: Please use commas to separate thousands for numbers with five or more digits (not for four digits) in the picture. e.g., "10000" should be "10,000"
%Author Will try
 to Figure \ref{fig1} for the main sequence  M-type 
($T_{\rm eff} = 3500$ K) and  
A-type stars ($T_{\rm eff} = 9800$). The continuum opacity is shown with dashed lines. The 
total (including lines) opacity is omitted for the solar case since it
is already shown in Figure \ref{fig1}. The representative data 
correspond to 
($\rho  \simeq 0.8 \times 10^{-9}$ g cm$^{-3}$, $T \simeq 8600$ K) in the case
of the A-type star and ($\rho  \simeq 1.8 \times 10^{-6}$ g cm$^{-3}$,
$T \simeq 3200$ K) for the M-type star. \label{fig3}}
\end{figure}

Figure \ref{fig3} repeats the information displayed in Figure \ref{fig1} 
for the Sun (shown here in orange), but also includes  a model for an 
 A-type star 
$T_{\rm eff} = 9800$ K) and a much cooler M-type star 
($T_{\rm eff} = 3500 $ K). As before, we have chosen atmospheric depths
representative of the regions where the optical continuum escapes.
The lower panel shows the photoionization (and bremsstrahlung) opacity
for the three stars with dashed lines. For the hot (A-type, blue) and cool
stars (M-type, red), the total opacity, including lines, is also shown.

The continuum of the M-type star is, like for the Sun, shaped by H$^-$
photoionization between 4000 and 16,000 \AA. H$^-$ free-free
opacity dominates at longer wavelengths, and the photoionization  of
heavier elements (and electron
scattering) is most relevant in the UV. However, 
line absorption severely blocks much of the continuum flux. 
The continuum of the warmer
A-type star, on the other hand, shows the characteristic shape of the 
hydrogenic photoionization cross-section, proportional to $\lambda^3$
[24], with discontinuities corresponding to the minimum ionization 
energies for each of the $n=1, 2, 3, \ldots$, etc. levels.
Note that the $H^-$ bremsstrahlung opacity shares the same slope [15].

Line absorption is progressively reduced as the surface temperature
of the star increases, due to the ionization of the main line absorbers,
chiefly atomic iron. It is quite obvious in Figure \ref{fig3} that the
lines add sharp opacity peaks on top of the continuum for the A-type star, while
the lower edge of the line absorption appears detached from the continuum
opacity for the cooler M-type star, indicative of a massive accumulation of
lines that overlap, mainly from molecules such as CH, CN, CO, OH, and TiO.

\section{Observations}
\label{observations}

A sanity check is always necessary, for which we may use the solar spectrum.
Nevertheless, it  turns out that performing observations of the  solar absolute flux
poses notable difficulties related to its overwhelming brightness, 
large angular size on the sky, and the hurdles to calibrate the observations 
using laboratory sources or sky sources such as standard stars. This is not to 
say that such observations have not been made. They, in fact, keep being made
with regularity, motivated by the need to understand the amount and variability
of the solar radiation impacting the Earth (see, e.g., [14,34]),  
 but to our knowledge, there is no solar spectrum of
reference available with a reliable calibration over a very broad spectral range.

An alternative is offered by the various solar analogs identified over the years,
largely with the motivation of performing differential determinations of 
chemical abundances, which benefit from the cancelation of systematic errors
inherent to that type of analysis. The star 18 Sco is the nearest such star available 
and has been studied in detail. Broad-coverage spectroscopy from the 
Hubble Space Telescope is available for this target, with high-quality 
absolute fluxes (see, e.g., [10]). 

\begin{figure}[H]
\includegraphics[width=14.5 cm]{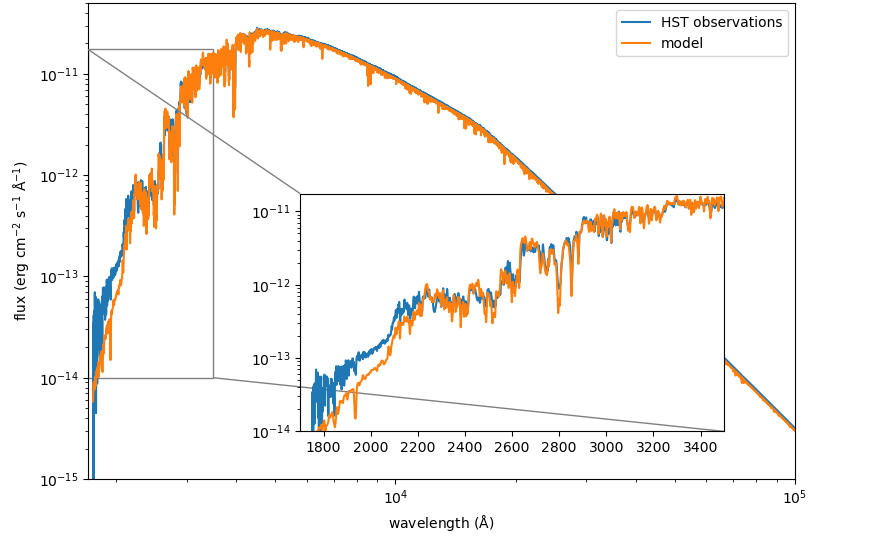}
\caption{Observed (blue) and model (orange) spectra for the solar twin
18 Sco. The model has been scaled using the determination of the 
star's angular diameter from [7].\label{fig4}}
\end{figure}

This star is close enough and bright enough that it has been 
observed with an interferometer, and its angular diameter has been resolved and measured to be
$\theta = 0.6759 \pm 0.0062$ milliarcseconds [7].
% or $0.663 \pm 0.007$ milliarseconds (Karovicova et al. 2022). 
The Gaia parallax for this star is  $70.7371 \pm 0.0631$ milliarcseconds, implying
it is at a distance $d$ of $14.14 \pm 0.01$ parsecs, and it has a radius of
$R = \theta / 2 \times d  
%= 0.6759 \times 10.^{-3} / 3600. / 180. *  \pi / 2. * 14.14 * 4.435 \times 10^7 
= 1.027 \pm 0.001 R_{\odot}$. The precision of the angular diameter, good to  
1 \%, allows us to scale the model fluxes, computed at the stellar surface, and
compare them to the flux measured for the star. Figure \ref{fig4} 
illustrates this exercise, using a solar model ($T_{\rm eff} = 5777$ K, 
$\log g = 4.437$, and the chemical composition from
 [2], [Fe/H]$=0$ ---very close to the parameters for 18 Sco published 
from [7]:
$T_{\rm eff} = 5817 \pm 4$ K, $\log g = 4.448 \pm 0.012$, and [Fe/H] $= 0.052 \pm 0.005$), 
showing  excellent agreement, except at the
shortest wavelengths ($\lambda < 2300 \AA$), where multiple assumptions built
in the models break down.

The light at wavelengths shorter than about 2000 \AA\ escapes from very high 
atmospheric layers, where the classical model atmospheres we have adopted are 
no longer realistic due mainly to 
departures from Local Thermodynamical Equilibrium, the breakdown of the assumption of 
hydrostatic equilibrium, and the relevance of magnetic fields, which are ignored. 
An ad hoc chromospheric temperature increase was thought in the 1970s to solve the flux 
discrepancies found for the Sun at short wavelengths [35,36], 
but hydrostatic models  cannot explain many of the observations 
such as the strength of CO bands [3].

A couple of features are noteworthy in Figure \ref{fig4}. At about 2500 \AA\, a break is
apparent, and a second one is visible near 2100 \AA. These sudden reductions in 
 flux correspond to the abrupt increases in opacity due to atomic 
magnesium and atomic aluminum, respectively. Other features from photoionization
edges one may expect based on Figure \ref{fig2} are not really visible, most likely,
as pointed out above, due to the opacity enhancement shifting the region  
from which the continuum flux escapes to layers higher up 
in the atmosphere. The continuum changes in slope at  
about 4000 \AA, longwards of which H$^-$ photoionization dominates, 
and 16,000 \AA, where H$^-$ bremsstrahlung becomes the main contributor
to the continuum opacity.

\begin{figure}[H]
\includegraphics[width=14.5 cm]{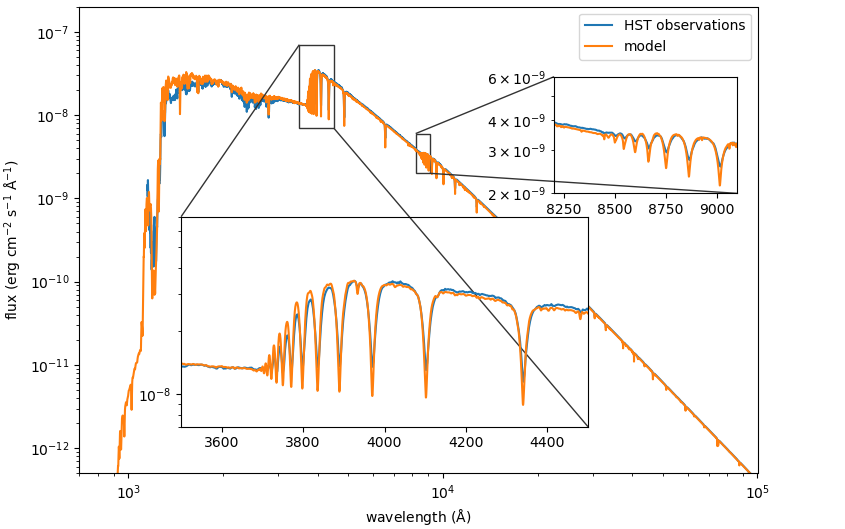}
\caption{Observed (blue) and model (orange) spectra for the A-type 
Sirius A. The model has been scaled using the determination of the 
star angular diameter by [13].\label{fig5}}
\end{figure}

The brightest star in the sky, Sirius, is a good example of an A-type star, and
its spectrum is included among the high-quality observations from the 
Hubble Space Telescope [9]. The star has a white dwarf companion that 
is irrelevant to our discussion. Its parallax is $379.21 \pm 1.58$ milliarseconds,
or a distance of 2.637 pc. The angular diameter has been measured by 
[23] to be $\theta = 6.039 \pm 0.019$ milliarcsends, and
more recently confirmed by [13] as 
$\theta = 6.041 \pm 0.017$ milliarconds). Figure~\ref{fig5} shows the observed
spectrum (blue) confronted with a model (orange) with $T_{\rm eff}$ = 10,000~K, 
$\log g = 4.0$ and (Fe/H])$=0$ 
 ([13] adopted for this star $T_{\rm eff} = 9845$ K, 
$\log g = 4.25$ and [Fe/H] = $+0.5$, where the two latter parameters are inherited from
[12]).

The jumps in flux due to the photoionization of hydrogen are quite obvious at 
 about 900 \AA\ ($n=1$), 3700 \AA\ ($n=2$), and 8200 \AA\ ($n=3$), becoming progressively
weaker. The drop in flux at about 1300 \AA\ is due to carbon photoionization, and
there is a smaller drop at about 1500 \AA\ caused by Si photoionization.
The very strong lines at about 1200 \AA\ are a blend of L$\alpha$ 
(H $n=2$ to $n=1$ transition) with S I lines on the blue side. The cores
of the H lines are deeper in the models than in the data, which could be 
a limitation in the models, although departures from Local Thermodynamic Equilibrium
work in the opposite sense and  can therefore be excluded (e.g., [33]), 
or an issue with scattered light in the observations.

As an example of a cool star, we have chosen Gliese 555, a well-studied red
dwarf star with an effective temperature of about 3200 K. This star is one of
the few M dwarfs included in the Hubble Space Telescope sample with accurate
fluxes, but unfortunately is not among the short list of M dwarfs with measured
angular diameters. Nonetheless, [27] have built a relation between
the infrared luminosity of M dwarfs and their radii, using the stars with 
interferometric angular diameters, and arrived at $R = 0.310 \pm 0.013 R_{\odot}$
for GJ 555, which, combined with the Gaia parallax, leads to $\theta = 0.461 \pm 0.019$ 
milliarcseconds. 

\begin{figure}[H]
\includegraphics[width=14.5 cm]{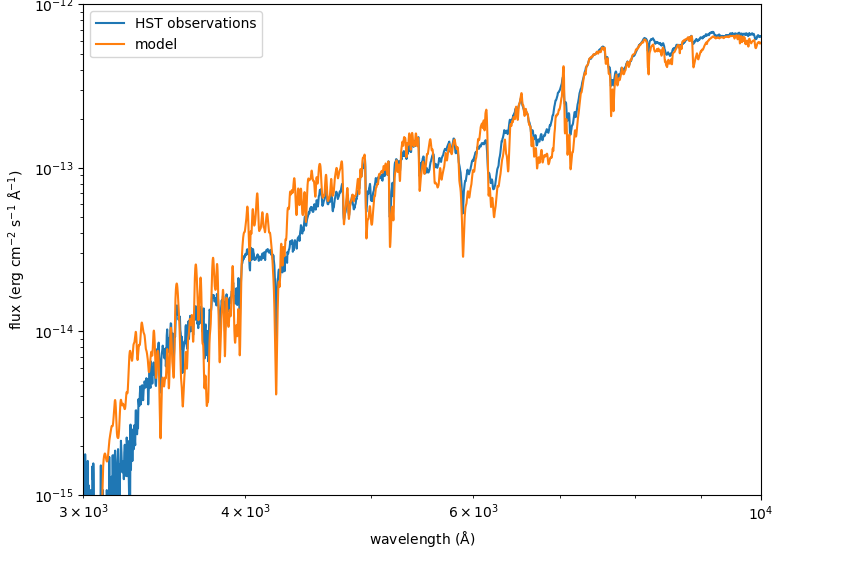}
\caption{Observed (blue) and model (orange) spectra for the M-type 
dwarf star GJ 555. The model has been scaled using the estimated 
stellar radius from [27] and the Gaia DR3 parallax for 
the star.\label{fig6}}
\end{figure}

Figure \ref{fig6} compares the observations with a model for $T_{\rm eff} = 3200$ K, 
$\log g = 5$ and [Fe/H]$=0$ ([27] give  $T_{\rm eff} = 3211$ K, 
$\log g = 4.89$ and [Fe/H]$=+0.17$), showing fair agreement. The parameters appear to be
appropriate, and so is the angular diameter, but the model's imperfections are
 much more significant than in the cases of 18 Sco and Sirius. The complexity
of the model is significantly higher due to the pervasive presence of molecules 
in the atmosphere of this star.
As one would expect from the analysis in \mbox{Section \ref{opacity},} and in particular
from Figure \ref{fig3}, the shape of the spectra of this type of star is
dominated by the presence of molecular bands of MgH, TiO, VO, and CaH, as well as
very strong lines from low-lying levels of Na I (5900 \AA\ and 8190 \AA), 
K I (7680 \AA), and Ca I (4227~\AA). 

%%%%%%%%%%%%%%%%%%%%%%%%%%%%%%%%%%%%%%%%%%
\section{Diagnostics}
\label{diagnostics}

Most relevant for the study of stars and understanding their properties is how 
opacity in general and photoionization in particular  vary depending upon
the properties of the star. 

The fundamental stellar parameters are mass
and radius, plus age and chemical composition. To the zeroth-order, mass determines
the fate of a star, including how long  its life will be, what chemical elements it will be able
to produce in its interior by nuclear fusion, and how it will  die, with the 
most typical outcomes being a core-collapse supernova leaving behind a
black hole or a neutron star or nothing for $M > 8 M_{\odot}$, or a white
dwarf for lower-mass stars. 
Nevertheless, from the point of view of the spectrum of a star, the most
relevant parameters are the star's surface (or {\it effective}) temperature, 
its surface gravity, and its chemical composition. 

In the longest phase of the life of a star, the {\it main sequence}, 
it fuses hydrogen in its core to produce helium. In this period, 
the mass correlates
perfectly with surface temperature: the more massive the star, the warmer the surface 
temperature. This is therefore the main atmospheric parameter that controls 
how the spectrum of the star looks, as illustrated in Figures \ref{fig4}--\ref{fig6}.  we now examine the impact of the other two parameters  on the 
structure of the stellar atmosphere and the shape of the spectrum.

I have already discussed the situation for cool, M-type stars, where 
H$^-$ photoionization is still the main contributor to the continuum
opacity in the optical and near-infrared, but the line absorption due to molecules 
dominates the opacity. In what follows, we will discuss the other two warmer
cases considered in our previous examples: a solar-like star and an A-type star. 

\begin{figure}[H]
\includegraphics[width=13 cm]{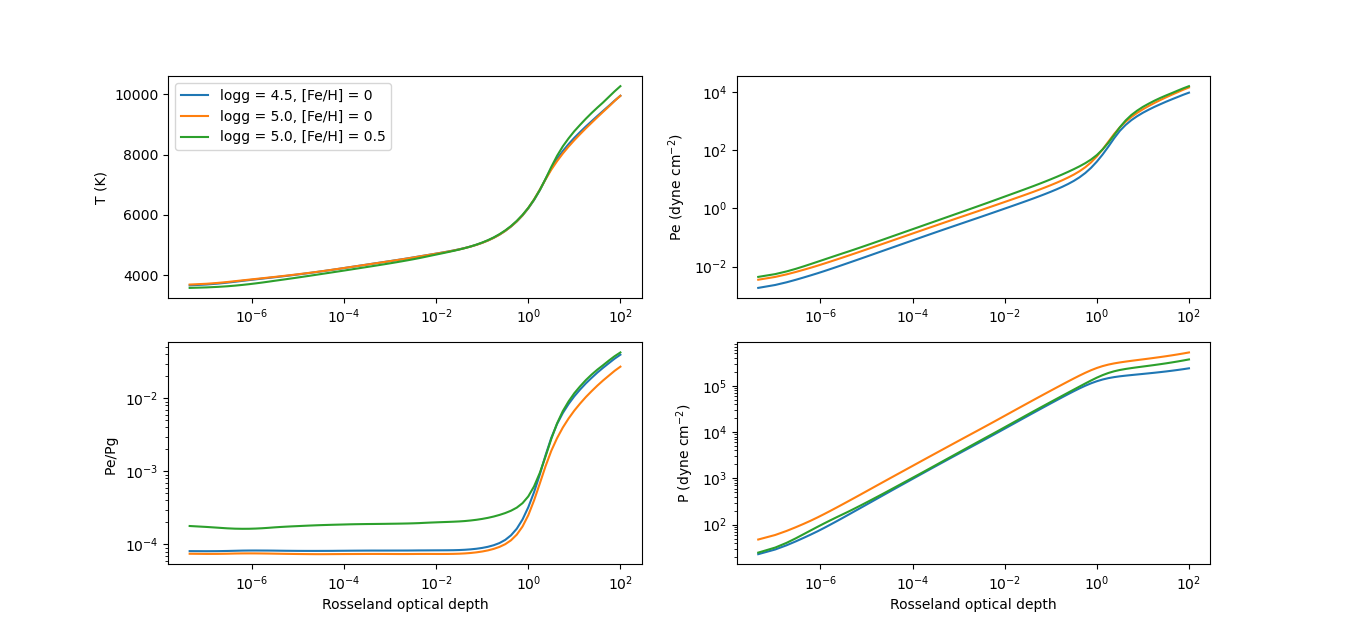}
\includegraphics[width=12 cm]{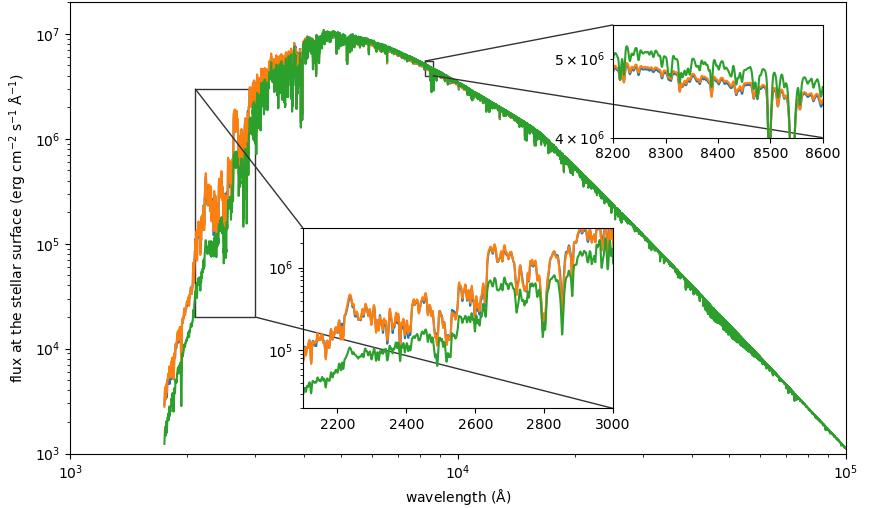}
\caption{(\textbf{Upper panel}) Run %MDPI: Please use commas to separate thousands for numbers with five or more digits (not for four digits) in the picture. e.g., "10000" should be "10,000"
%Author Will try
 of various thermodynamical quantities 
(clockwise: temperature, electron pressure ($Pe$), gas pressure ($P$) 
and their ratio ($Pe/P$)) 
with the Rosseland mean opacity for model atmospheres for a 
solar-like star ($T_{\rm eff}$ = 5777 K). 
(\textbf{Lower panel}) Predicted emergent flux at the atmospheric surface 
for the models in the upper panels (no scaling is necessary, since we are only 
comparing models). \label{fig7}}
\end{figure}

In the top panel of Figure \ref{fig7}, we can see the 
run of the main thermodynamical quantities for a solar-like star 
with Rosseland optical depth, which 
is a weighted mean of the integrated opacity along the atmosphere down to
a given depth, and gives a very useful reference axis when studying 
optical properties. There are three
models shown in the figure: a reference solar-like model (in blue), another with 0.5 dex higher
gravity (orange), and a third, which, in addition to the higher gravity, has
a higher metal content ([Fe/H] = $+0.5$). The bottom panel
of the figure shows the corresponding model spectra.

Our models assume that the atmosphere is in hydrostatic equilibrium
\begin{equation}
\frac{dP}{dm} = g,
\end{equation}
where $P$ is the gas pressure (turbulent and radiation pressure are negligible
for this type of atmosphere), $m$ is the mass colum, and $g$ is the gravitational
acceleration $g = G M/R^2$. An increase in gravity compresses the atmosphere, 
enhancing the pressure, while keeping the fractional contribution 
from electrons ($Pe$) to  it at a similar level ($10^{-4}$, except in the deepest
layers where H begins to be ionized). The effect of this change on the continuum
opacity is negligible, since the abundance of H$^-$, proportional to 
$Pe$, increases only mildy, while $Pe/P$ stays nearly constant.

On the other hand, an additional increase 
in the abundance of the heavy elements has a profound impact on the near-UV opacity, 
 due to the importance of iron and magnesium photoionization, and the increase in iron 
(and other metal) lines, 
and the UV flux is consequently reduced. The enhanced UV opacity cools down the outer
atmospheric layers and heats the deepest ones (an effect known as {\it backwarming}). 
There is a small increase in the 
electron  pressure, partly from the increase in the abundance of electron donors 
such as sodium, magnesium, calcium, and iron, which 
would in principle boost the optical and near-IR opacity, 
therefore reducing  the flux at those wavelengths, 
but the effect observed is exactly the opposite. This is in part due to the fact
that the relative enhancement of electron pressure in high atmospheric layers
disappears when reaching the continuum-forming layers at the Rosseland 
optical depth near unity. Furthermore, an increase in continuum opacity does not
necessarily imply a reduction in flux, since the model needs to self-adjust 
to satisfy energy conservation, which for these three models imposes that the
flux integrated over all wavelengths must be the same 
\begin{equation}
\int_0^{\infty} F_{\lambda} d\lambda = F = \sigma T_{\rm eff}^4,
\end{equation}
and therefore a decrease 
in the UV flux has to be compensated at other wavelengths.

\begin{figure}[H]
\includegraphics[width=13 cm]{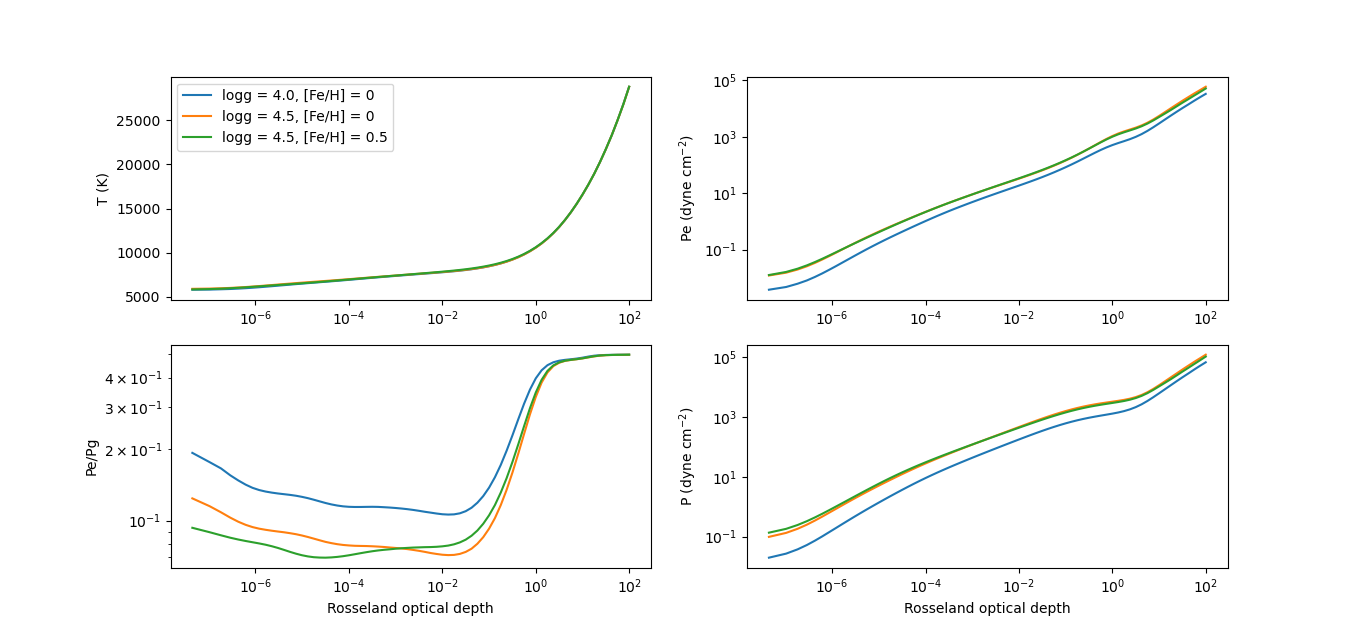}
\includegraphics[width=12 cm]{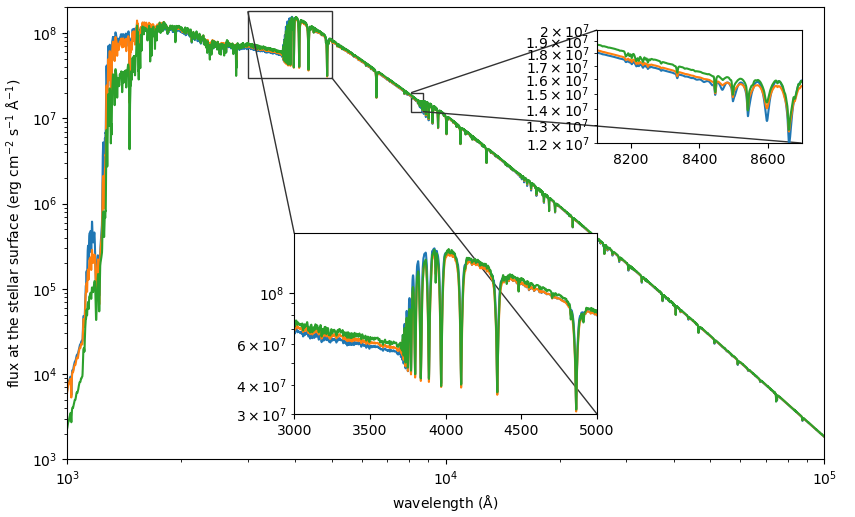}
\caption{(\textbf{Upper panel}) Run %MDPI: Please use commas to separate thousands for numbers with five or more digits (not for four digits) in the picture. e.g., "10000" should be "10,000"
%Author Will try
 of various thermodynamical quantities 
(clockwise: temperature, electron pressure ($Pe$), gas pressure ($P$), 
and their ratio ($Pe/P$)) 
with the Rosseland mean opacity for model atmospheres for an A-type star 
($T_{\rm eff}$ = 9750 K). 
(\textbf{Lower panel}) Predicted emergent flux at the atmospheric surface 
for the models in the upper panels.\label{fig8}}
\end{figure}

The situation for an A-type star such as Sirius, illustrated in Figure \ref{fig8},
is different. Here, H$^-$ plays only a minor role in the continuum opacity, 
and hydrogen atoms are the main contributors in the optical and
near-infrared regions. An increase in surface gravity leaves the run of temperature
with optical depth unchanged but compresses the atmosphere, enhancing the
gas pressure and, to a lesser extent, the electron density, with an overall 
reduction in the electron's partial pressure.  An  additional increase in 
the abundances of the heavy elements does not change the atmospheric 
structure much.
%, although it must enhance slightly the mean molecular weight. 

In the lower panel of Figure \ref{fig8}, we can appreciate how the changes
described affect the emergent radiative flux. The boost in pressure associated 
with the increase in surface gravity broadens the lines somewhat , both the hydrogen
lines and other strong features. The pressure enhancement and the subsequent 
reduction in the electron partial pressure increases the ionization 
and dampens slightly the hydrogen photoionization, 
as clearly visible in the Balmer (3700 \AA) and  the Paschem (8500 \AA) jumps.
An enhancement in the metal abundance noticeably affects  the UV absorption, in 
this case mainly due to (ionized) iron photoionization, reducing the UV flux, 
which is compensated with a slightly increase at other wavelengths to keep 
the integrated flux constant. 

%%%%%%%%%%%%%%%%%%%%%%%%%%%%%%%%%%%%%%%%%%
\section{Summary and Conclusions}
\label{conclusions}

We have used a state-of-the-art code for computing synthetic spectra and 
standard plane-parallel model atmospheres to evaluate the role of 
photoionization in shaping the spectral energy distributions of stars.

The photoionization of atomic hydrogen or the H$^-$ ion are a  dominant 
source of opacity in the optical and infrared regions in stars with  
one  solar mass  or more. While H$^-$ remains 
a dominant contributor to the optical/infrared continuum opacity, 
molecular line opacity  becomes more important in cooler (usually 
less massive) stars 
and causes  a significant redistribution of the emergent flux. 
This makes it harder to model stellar spectra of M-type dwarfs, 
which are the most common stars across the Milky Way, than warmer stars.

The ultraviolet spectra of most stars are dominated by photoionization
from heavier elements (magnesium, aluminum, silicon, and iron), as well
as atomic line absorption. In models of solar-type stars, the ultraviolet 
opacity seems to be accounted for appropriately, although a more exhaustive
study is necessary to make sure that (a) no important contributors are 
missed (e.g., photoionization from elements with similar atomic mass or 
heavier than iron), and (b) no opacity sources are  counted twice, 
(e.g., autoionization lines being included in the atomic line list and
 in the photoionization cross-sections).
 
The illustrative examples given in this paper can serve as a starting 
point for new, deeper, investigations, looking at the sources of
opacity in various types of stars. Improving our understanding of 
the atmospheric opacity paves the way to refining  the agreement  
between the observed spectral energy distributions of stars and model
predictions, which are key to our ability to infer stellar properties, 
such as mass, radius, luminosity, chemical composition, etc., 
from observations.

%%%%%%%%%%%%%%%%%%%%%%%%%%%%%%%%%%%%%%%%%%
%\section{Patents}

%This section is not mandatory, but may be added if there are patents resulting from the work reported in this manuscript.

%%%%%%%%%%%%%%%%%%%%%%%%%%%%%%%%%%%%%%%%%%
\vspace{6pt} 

%%%%%%%%%%%%%%%%%%%%%%%%%%%%%%%%%%%%%%%%%%
%% optional
%\supplementary{The following supporting information can be downloaded at:  \linksupplementary{s1}, Figure S1: title; Table S1: title; Video S1: title.}


%%%%%%%%%%%%%%%%%%%%%%%%%%%%%%%%%%%%%%%%%%
%\authorcontributions{For research articles with several authors, a short paragraph specifying their individual contributions must be provided. The following statements should be used ``Conceptualization, X.X. and Y.Y.; methodology, X.X.; software, X.X.; validation, X.X., Y.Y. and Z.Z.; formal analysis, X.X.; investigation, X.X.; resources, X.X.; data curation, X.X.; writing---original draft preparation, X.X.; writing---review and editing, X.X.; visualization, X.X.; supervision, X.X.; project administration, X.X.; funding acquisition, Y.Y. All authors have read and agreed to the published version of the manuscript.'', please turn to the  \href{http://img.mdpi.org/data/contributor-role-instruction.pdf}{CRediT taxonomy} for the term explanation. Authorship must be limited to those who have contributed substantially to the work~reported.}

\funding{The author acknowledges support for this research from the 
Spanish Ministry of Science and Innovation (MICINN) 
projects s AYA2017-86389-P and PID2020-117493GBI00. Funding for the DPAC has been provided by national institutions, in particular the institutions
participating in the {\it Gaia} Multilateral Agreement.}

%\institutionalreview{}

%\informedconsent{Any research article describing a study involving humans should contain this statement. Please add ``Informed consent was obtained from all subjects involved in the study.'' OR ``Patient consent was waived due to REASON (please provide a detailed justification).'' OR ``Not applicable'' for studies not involving humans. You might also choose to exclude this statement if the study did not involve humans.

%Written informed consent for publication must be obtained from participating patients who can be identified (including by the patients themselves). Please state ``Written informed consent has been obtained from the patient(s) to publish this paper'' if applicable.}

\institutionalreview{Not applicable}
%{In this section, you should add the Institutional Review Board Statement and approval number, if relevant to your study. You might choose to exclude this statement if the study did not require ethical approval. Please note that the Editorial Office might ask you for further information. Please add “The study was conducted in accordance with the Declaration of Helsinki, and approved by the Institutional Review Board (or Ethics Committee) of NAME OF INSTITUTE (protocol code XXX and date of approval).” for studies involving humans. OR “The animal study protocol was approved by the Institutional Review Board (or Ethics Committee) of NAME OF INSTITUTE (protocol code XXX and date of approval).” for studies involving animals. OR “Ethical review and approval were waived for this study due to REASON (please provide a detailed justification).” OR “Not applicable” for studies not involving humans or animals.}

\informedconsent{Not applicable}
%{Any research article describing a study involving humans should contain this statement. Please add ``Informed consent was obtained from all subjects involved in the study.'' OR ``Patient consent was waived due to REASON (please provide a detailed justification).'' OR ``Not applicable'' for studies not involving humans. You might also choose to exclude this statement if the study did not involve humans. Written informed consent for publication must be obtained from participating patients who can be identified (including by the patients themselves). Please state ``Written informed consent has been obtained from the patient(s) to publish this paper'' if applicable.}

\dataavailability{The opacity tables and the 
model spectra used in this paper have been computed with version v1.2 
of Synple, publicly available from github.com/callendeprieto/synple.}

\acknowledgments{I am thankful to Ivan Hubeny for his comments 
on an early draft of this manuscript. This work has made use of data from the European Space Agency (ESA) mission
{\it Gaia} (\url{https://www.cosmos.esa.int/gaia}), processed by the {\it Gaia}
Data Processing and Analysis Consortium (DPAC,
\url{https://www.cosmos.esa.int/web/gaia/dpac/consortium}). This research has made use of the SIMBAD database, operated at CDS, Strasbourg, France.
 } 

\conflictsofinterest{The author  declares no conflicts of interest.}
%Authors must identify and declare any personal circumstances or interest that may be perceived as inappropriately influencing the representation or interpretation of reported research results. Any role of the funders in the design of the study; in the collection, analyses or interpretation of data; in the writing of the manuscript; or in the decision to publish the results must be declared in this section. If there is no role, please state ``The funders had no role in the design of the study; in the collection, analyses, or interpretation of data; in the writing of the manuscript; or in the decision to publish the~results''.} 

%%%%%%%%%%%%%%%%%%%%%%%%%%%%%%%%%%%%%%%%%%
%% Optional
%\sampleavailability{Samples of the compounds ... are available from the authors.}

%% Only for journal Encyclopedia
%\entrylink{The Link to this entry published on the encyclopedia platform.}

%\abbreviations{Abbreviations}{
%The following abbreviations are used in this manuscript:\\

%\noindent 
%\begin{tabular}{@{}ll}
%MDPI & Multidisciplinary Digital Publishing Institute\\
%DOAJ & Directory of open access journals\\
%TLA & Three letter acronym\\
%LD & Linear dichroism
%\end{tabular}
%}

%%%%%%%%%%%%%%%%%%%%%%%%%%%%%%%%%%%%%%%%%%
%% Optional

%%%%%%%%%%%%%%%%%%%%%%%%%%%%%%%%%%%%%%%%%%
\begin{adjustwidth}{-\extralength}{0cm}
\printendnotes[custom]

\reftitle{References}

\PublishersNote{}
\end{adjustwidth}
\end{document}